\documentclass[10pt]{article}

\pdfoutput=1

\usepackage{graphicx, color}
\usepackage{booktabs}
\usepackage{subfigure}
\usepackage{amsmath}
\usepackage[margin=1in]{geometry}
\usepackage{listings}
\usepackage{url}
\usepackage{comment}
\usepackage{fancyhdr}
\usepackage{setspace}

\usepackage{cite}
\usepackage{blkarray} %
\usepackage{multirow} %

\newcommand{\planycrespondents}{1,436}
\newcommand{\planycresponses}{31,893}
\newcommand{\planycseedideas}{25}
\newcommand{\planycuploadedideas}{464}
\newcommand{\planycactiveideas}{269}
\newcommand{\planycuploadedactiveideas}{244}

\newcommand{\planyctoptenuploaded}{8}
\newcommand{\oecdrespondents}{1,668}
\newcommand{\oecdresponses}{28,852}
\newcommand{\oecdseedideas}{60}
\newcommand{\oecduploadedideas}{534}
\newcommand{\oecdactiveideas}{285}
\newcommand{\oecduploadedactiveideas}{231}
\newcommand{\oecdseedideasdeactivated}{6}
\newcommand{\oecdtoptenuploaded}{7}

\newcommand{\xdot}{90\% } %

\begin{document}

\title{Wiki surveys:\\Open and quantifiable social data collection\footnotemark[1]}
\author{Matthew J. Salganik\footnotemark[2]~~and Karen E.C. Levy\footnotemark[3]}
\date{}

\renewcommand{\thefootnote}{\fnsymbol{footnote}}
\thispagestyle{empty}
\footnotetext[1]{We thank Peter Lubell-Doughtie, Adam Sanders, Pius Uzamere, Dhruv Kapadia, Chap Ambrose, Calvin Lee, Dmitri Garbuzov, Brian Tubergen, Peter Green, and Luke Baker for outstanding web development; we thank Nadia Heninger, Bill Zeller, Bambi Tsui, Dhwani Shah, Gary Fine, Mark Newman, Dennis Feehan, Sophia Li, Lauren Senesac, Devah Pager, Paul DiMaggio, Adam Slez, Scott Lynch, David Rothschild, and Ceren Budak for valuable suggestions; and we thank Josh Weinstein for his critical role in the genesis of this project.  Further, we thank Ibrahim Abdul-Matin and colleagues at the New York City Mayor's Office and Joanne Caddy, Julie Harris, and Cassandra Davis at the Organisation for Economic Co-operation and Development.  This research was supported by grants from Google (Faculty Research Award, the People and Innovation Lab, and Summer of Code 2010); Princeton University Center for Information Technology Policy; Princeton University Committee on Research in the Humanities and Social Sciences; the National Science Foundation [grant number CNS-0905086]; and the National Institutes of Health [grant number P32-CHD047879].   Some of this research was performed while MJS was employed by Microsoft Research.  This paper represents the views of its authors and not the users or funders of \url{www.allourideas.org}.  All data collection and protection procedures were approved by the Institutional Review Board of Princeton University (protocol 4885).  Upon publication of this article, all data and code needed to reproduce these results will be deposited into the archive of the Office of Population Research at Princeton University.}

\footnotetext[2]{Department of Sociology, Center for Information Technology Policy, and Office of Population Research, Princeton University, Princeton, NJ, USA, \texttt{mjs3@princeton.edu}.}
\footnotetext[3]{Karen E.C. Levy, Information Law Institute and Department of Media, Culture, and Communication, New York University, New York, NY, USA and Data \& Society Research Institute, New York, NY, USA, \texttt{karen.levy@nyu.edu}.}
\renewcommand{\thefootnote}{\arabic{footnote}}

\maketitle

\begin{abstract}
In the social sciences, there is a longstanding tension between data collection methods that facilitate quantification and those that are open to unanticipated information.  Advances in technology now enable new, hybrid methods that combine some of the benefits of both approaches.  Drawing inspiration from online information aggregation systems like Wikipedia and from traditional survey research, we propose a new class of research instruments called \emph{wiki surveys}. Just as Wikipedia evolves over time based on contributions from participants, we envision an evolving survey driven by contributions from respondents. We develop three general principles that underlie wiki surveys: they should be greedy, collaborative, and adaptive. Building on these principles, we develop methods for data collection and data analysis for one type of wiki survey, a pairwise wiki survey. Using two proof-of-concept case studies involving our free and open-source website \url{www.allourideas.org}, we show that pairwise wiki surveys can yield insights that would be difficult to obtain with other methods.  
\end{abstract}

\newpage

\section{Introduction}

In the social sciences, there is a longstanding tension between data collection methods that facilitate quantification and those that are open to unanticipated information.  For example, one can contrast a traditional public opinion survey based on a series of pre-written questions and answers with an interview in which respondents are free to speak in their own words.  The tension between these approaches derives, in part, from the strengths of each: open approaches (e.g., interviews) enable us to learn new and unexpected information, while closed approaches (e.g., surveys) tend to be more cost-effective and easier to analyze.  Fortunately, advances in technology now enable new, hybrid approaches that combine the benefits of each.  Drawing inspiration both from online information aggregation systems like Wikipedia and from traditional survey research, we propose a new class of research instruments called \emph{wiki surveys}. Just as Wikipedia grows and improves over time based on contributions from participants, we envision an evolving survey driven by contributions from respondents.

Although the tension between open and closed approaches to data collection is currently most evident in disagreements between proponents of quantitative and qualitative methods, the trade-off between open and closed survey questions was also particularly contentious in the early days of survey research~\cite{lazarsfeld_controversy_1944, converse_strong_1984, converse_survey_2009}.  Although closed survey questions, in which respondents choose from a series of pre-written answer choices, have come to dominate the field, this is not because they have been proven superior for measurement.  Rather, the dominance of closed questions is largely based on practical considerations: having a fixed set of responses dramatically simplifies data analysis~\cite{schuman_method_2008}.  

The dominance of closed questions, however, has led to some missed opportunities, as open approaches may provide insights that closed methods cannot~\cite{schuman_open_1979, schuman_problems_1987, presser_measurement_1990, schuman_method_2008}.  For example, in one study, researchers conducted a split-ballot test of an open and closed form of a question about what people value in jobs~\cite{schuman_open_1979}.  When asked in closed form, virtually all respondents provided one of the five researcher-created answer choices.  But, when asked in open form, nearly 60\% of respondents provided a new answer that fell outside the original five choices.  In some situations, these unanticipated answers can be the most valuable, but they are not easily collected with closed questions.  Because respondents tend to confine their responses to the choices offered~\cite{krosnick_survey_1999}, researchers who construct all the possible choices necessarily constrain what can be learned.  

Projects that depend on crowdsourcing and user-generated content, such as Wikipedia, suggest an alternative approach.  What if a survey could be constructed by respondents themselves?  Such a survey could produce clear, quantifiable results at a reasonable cost, while minimizing the degree to which researchers must impose their pre-existing knowledge and biases on the data collection process. We see wiki surveys as an initial step toward this possibility.

Wiki surveys are intended to serve as a complement to, not a replacement for, traditional closed and open methods.  In some settings, traditional methods will be preferable, but in others we expect that wiki surveys may produce new insights. The field of survey research has always evolved in response to new opportunities created by changes in technology and society~\cite{mitofsky_presidential_1989, dillman_presidential_2002, couper_designing_2008, couper_web_2009, couper_future_2011, groves_three_2011, newport_presidential_2011}, and we see this research as part of that longstanding evolution.  

In this paper, we develop three general principles that underlie wiki surveys: they should be greedy, collaborative, and adaptive. Building on these principles, we develop methods for data collection and data analysis for one type of wiki survey, a pairwise wiki survey. Using two proof-of-concept case studies involving our free and open-source website \url{www.allourideas.org}, we show that pairwise wiki surveys can yield insights that would be difficult to obtain with other methods.  The paper concludes with a discussion of the limitations of this work and possibilities for future research.

\section{Wiki surveys}

Online information aggregation projects, of which Wikipedia is an exemplar, can inspire new directions in survey research. These projects, which are built from crowdsourced, user-generated content, tend to share certain properties that are not characteristic of traditional surveys~\cite{benkler_wealth_2006, howe_crowdsourcing:_2009, noveck_wiki_2009, nielsen_reinventing_2012}.  These properties guide our development of wiki surveys. In particular, we propose that wiki surveys should follow three general principles: they should be \emph{greedy}, \emph{collaborative}, and \emph{adaptive}. 

\subsection{Greediness}

\begin{figure}
\centering
\includegraphics[width=0.6\textwidth]{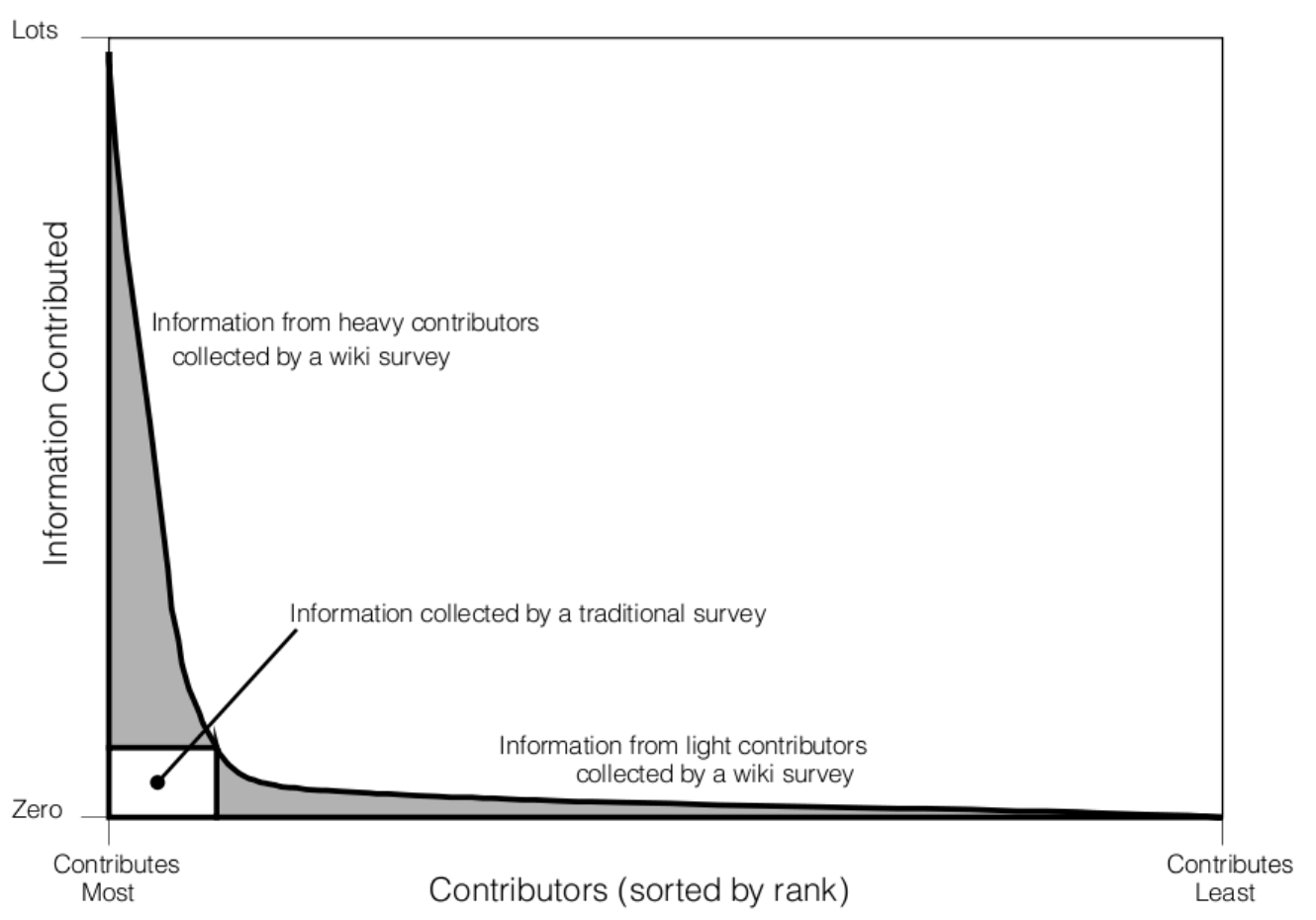}
\caption{{\bf Schematic of rank order plot of contributions to successful online information aggregation projects.}  These systems can handle both heavy contributors (``the fat head''), shown on the left side of the plot, and light contributors (``the long tail''), shows on the right side of the plot.  Traditional survey methods utilize information from neither the ``fat head'' nor the ``long tail'' and thus leave huge amounts of information uncollected.}
\label{fig:fathead_longtail}
\end{figure}

Traditional surveys attempt to collect a fixed amount of information from each respondent; respondents who want to contribute less than one questionnaire's worth of information are considered problematic, and respondents who want to contribute more are prohibited from doing so.  This contrasts sharply with successful information aggregation projects on the Internet, which collect as much or as little information as each participant is willing to provide.  Such a structure typically results in highly unequal levels of contribution: when contributors are plotted in rank order, the distributions tend to show a small number of heavy contributors---the ``fat head''---and a large number of light contributors---the ``long tail''~\cite{anderson_long_2006, wilkinson_strong_2008} (Fig.~\ref{fig:fathead_longtail}). For example, the number of edits to Wikipedia per editor roughly follows a power-law distribution with an exponent 2~\cite{wilkinson_strong_2008}.  If Wikipedia were to allow 10 and only 10 edits per editor---akin to a survey that requires respondents to complete one and only one form---it would exclude about 95\% of the edits contributed.   As such, traditional surveys potentially leave enormous amounts of information from the ``fat head'' and ``long tail'' uncollected.  Wiki surveys, then, should be \emph{greedy} in the sense that they should capture as much or as little information as a respondent is willing to provide.  

\subsection{Collaborativeness}

In traditional surveys, the questions and answer choices are typically written by researchers rather than respondents.  In contrast, wiki surveys should be \emph{collaborative} in that they are open to new information contributed directly by respondents that may not have been anticipated by the researcher, as often happens during an interview. Crucially, unlike a traditional ``other" box in a survey, this new information would then be presented to future respondents for evaluation.  In this way, a wiki survey bears some resemblance to a focus group in which participants can respond to the contributions of others~\cite{merton_focused_1946, merton_focussed_1987}.  Thus, just as a community collaboratively writes and edits Wikipedia, the content of a wiki survey should be partially created by its respondents.  This approach to collaborative survey construction resembles some forms of survey pre-testing~\cite{presser_methods_2004}.  However, rather than thinking of pre-testing as a phase distinct from the actual data collection, in wiki surveys the collaboration process continues throughout data collection.

\subsection{Adaptivity}

Traditional surveys are static: survey questions, their order, and their possible answers are determined before data collection begins and do not evolve as more is learned about the parameters of interest.  This static approach, while easier to implement, does not maximize the amount that can be learned from each respondent.  Wiki surveys, therefore, should be \emph{adaptive} in the sense that the instrument is continually optimized to elicit the most useful information, given what is already known.  In other words, while collaborativeness involves being open to new information, adaptivity involves using the information that has already been gathered more efficiently.  In the context of wiki surveys, adaptivity is particularly important given that respondents can provide different amounts of information (due to greediness) and that some answer choices are newer than others (due to collaborativeness).  Like greediness and collaborativeness, adaptivity increases the complexity of data analysis.  However, research in related areas~\cite{balasubramanian_measuring_1989, singh_adaptive_1990, groves_responsive_2006, toubia_adaptive_2007, smyth_open-ended_2009, chen_usher:_2010, dzyabura_active_2011, montgomery_computerized_2013} suggests that gains in efficiency from adaptivity can more than offset the cost of added complexity.  

\section{Pairwise Wiki Surveys}

Building on previous work~\cite{lewry_kittenwar:_2007, wu_usg_2008, weinstein_photocracy:_2009, shah_solving_2009, das_sarma_ranking_2010, luon_rankr:_2012, salesses_collaborative_2013}, we operationalize these three principles into what we call a \emph{pairwise wiki survey}.  A pairwise wiki survey consists of a single question with many possible answer items.  Respondents can participate in a pairwise wiki survey in two ways: first, they can make pairwise comparisons between items (i.e., respondents vote between item A and item B), and second, they can add new items that are then presented to future respondents.

Pairwise comparison, which has a long history in the social sciences \cite{thurstone_method_1927}, is an ideal question format for wiki surveys because it is amenable to the three criteria described above.  Pairwise comparison can be \emph{greedy} because the instrument can easily present as many (or as few) prompts as each respondent is willing to answer.  New items contributed by respondents can easily be integrated into the choice sets of future respondents, enabling the instrument to be \emph{collaborative}.  Finally, pairwise comparison can be \emph{adaptive} because the pairs to be presented can be selected to maximize learning given previous responses. These properties exist because pairwise comparisons are both granular and modular; that is, the unit of contribution is small and can be readily aggregated~\cite{benkler_wealth_2006}. 

Pairwise comparison also has several practical benefits. First, pairwise comparison makes manipulation, or ``gaming," of results difficult because respondents cannot choose which pairs they will see; instead, this choice is made by the instrument. Thus, when there is a large number of possible items, a respondent would have to respond many times in order to be presented with the item that she wishes to ``vote up" (or ``vote down")~\cite{hacker_matchin:_2009}.  Second, pairwise comparison requires respondents to prioritize items---that is, because the respondent must select one of two discrete answer choices from each pair, she is prevented from simply saying that she likes (or dislikes) every option equally strongly.  This feature is particularly valuable in policy and planning contexts, in which finite resources make prioritization of ideas necessary. Finally, responding to a series of pairwise comparisons is reasonably enjoyable, a common characteristic of many successful web-based social research projects~\cite{salganik_web-based_2009, goel_real_2010}.  

\subsection{Data collection}

In order to collect pairwise wiki survey data, we created the free and open-source website All Our Ideas (\url{www.allourideas.org}), which enables anyone to create their own pairwise wiki survey. To date, about 5,000 pairwise wiki surveys have been created that include about 200,000 items and 5 million responses.  By providing this service online, we are able to collect a tremendous amount of data about how pairwise wiki surveys work in practice, and our steady stream of users provides a natural testbed for further methodological research.  

The data collection process in a pairwise wiki survey is illustrated by a project conducted by the New York City Mayor's Office of Long-Term Planning and Sustainability in order to integrate residents' ideas into PlaNYC 2030, New York's citywide sustainability plan.  The City has typically held public meetings and small focus groups to obtain feedback from the public.  By using a pairwise wiki survey, the Mayor's Office sought to broaden the dialogue to include input from residents who do not traditionally attend public meetings. To begin the process, the Mayor's Office generated a list of 25 ideas based on their previous outreach (e.g., ``Require all big buildings to make certain energy efficiency upgrades," ``Teach kids about green issues as part of school curriculum").  

\begin{figure}
  \includegraphics[width=\textwidth]{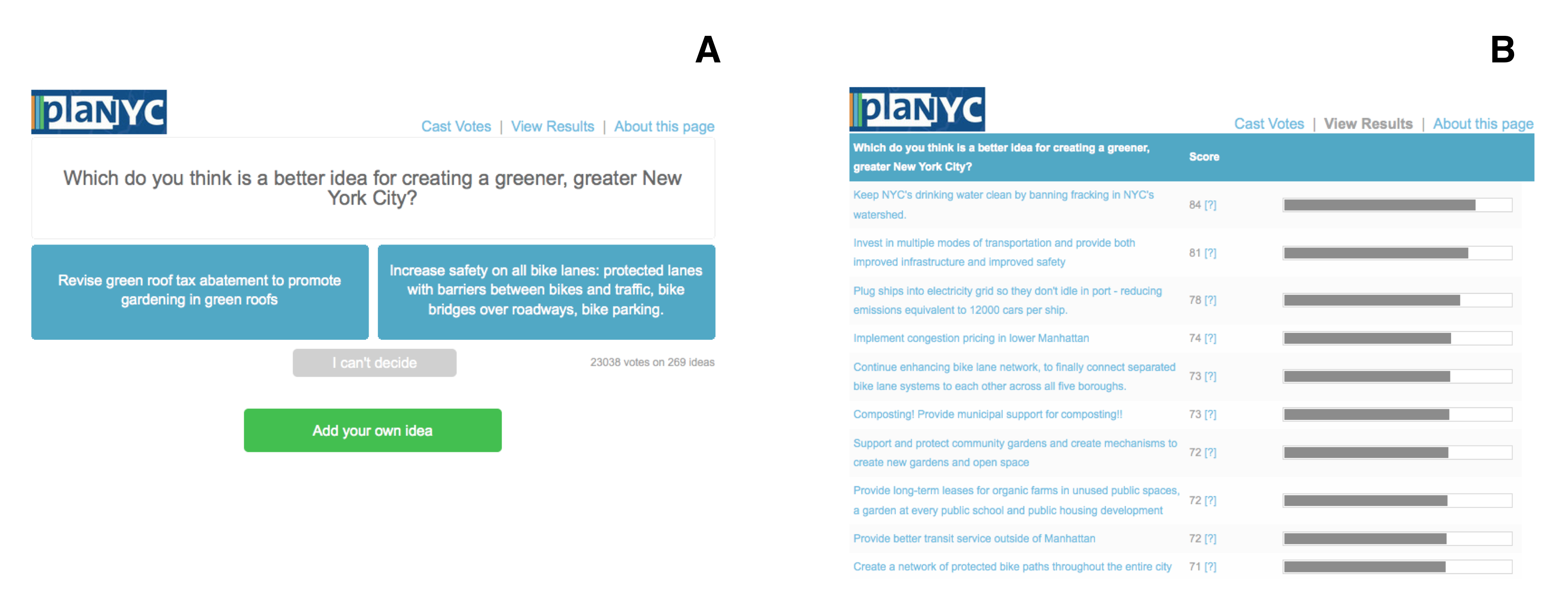}
  \centering
  \caption{{\bf Response and results interfaces at \protect \url{www.allourideas.org}.}  This example is from a pairwise wiki survey created by the New York City Mayor's Office to learn about residents' ideas about how to make New York ``greener and greater.''}
     \label{fig:screenshots} %
\end{figure}

Using these 25 ideas as ``seeds," the Mayor's Office created a pairwise wiki survey with the question ``Which do you think is a better idea for creating a greener, greater New York City?" Respondents were presented with a pair of ideas (e.g., ``Open schoolyards across the city as public playgrounds" and ``Increase targeted tree plantings in neighborhoods with high asthma rates"), and asked to choose between them (see Fig.~\ref{fig:screenshots}).  After choosing, respondents were immediately presented with another randomly selected pair of ideas (the process for choosing the pairs is described in SI 1).  Respondents were able to continue contributing information about their preferences for as long as they wished by either voting or choosing ``I can't decide."  Crucially, at any point, respondents were able to contribute their own ideas, which---pending approval by the wiki survey creator---became part of the pool of ideas to be presented to others.  Respondents were also able to view the popularity of the ideas at any time, making the process transparent.  However, by decoupling the processes of voting and viewing the results---which occur on distinct screens (see Fig.~\ref{fig:screenshots})---the site prevents a respondent from having immediate information about the opinions of others when she responds, which minimizes the risk of social influence and information cascades~\cite{salganik_experimental_2006, salganik_web-based_2009, zhu_switch_2012, muchnik_social_2013, van_de_rijt_field_2014}.

The Mayor's Office launched its pairwise wiki survey in October 2010 in conjunction with a series of community meetings to obtain resident feedback. The effort was publicized at meetings in all five boroughs of the city and via social media.  Over about four months, \planycrespondents~respondents contributed \planycresponses~responses and \planycuploadedideas~ideas to the pairwise wiki survey.

\subsection{Data analysis}

\begin{figure}
\includegraphics[width=\textwidth]{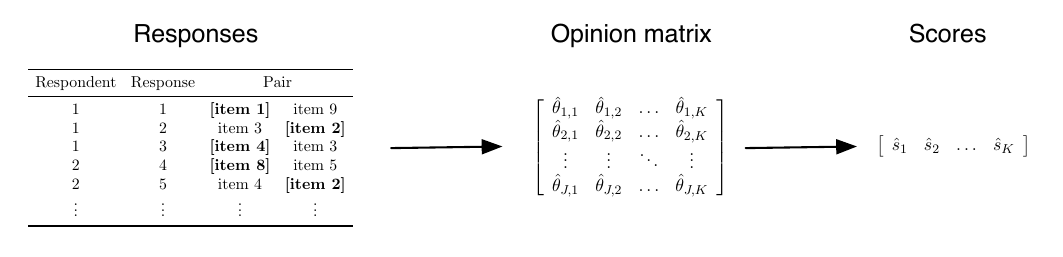}
\centering
\caption{{\bf Summary of data analysis plan.}  We use responses to estimate the opinion matrix $\mbox{\boldmath$\Theta$}$ and then we summarize the opinion matrix with the scores of each item.}
\label{fig:analysis_summary}
\end{figure}

Given this data collection process, we analyze data from a pairwise wiki survey in two main steps (Fig.~\ref{fig:analysis_summary}).  First, we use responses to estimate the opinion matrix $\mbox{\boldmath$\Theta$}$ that includes an estimate of how much each respondent values each item. Next, we summarize the opinion matrix to produce a score for each item that estimates the probability that it will beat a randomly chosen item for a randomly chosen respondent.  Because this analysis is modular, either step---estimation or summarization---could be improved independently.

\subsubsection{Estimating the opinion matrix}

The analysis begins with a set of pairwise comparison responses that are nested within respondents.  For example, Fig.~\ref{fig:analysis_summary} shows five hypothetical responses from two respondents.  These responses are used to estimate the opinion matrix 
\begin{equation*}
\mbox{\boldmath$\Theta$} =
\left[
\begin{array}{cccc}
\theta_{1,1} & \theta_{1,2} & \ldots & \theta_{1,K} \\
\theta_{2,1} & \theta_{2,2} & \ldots & \theta_{2,K} \\
\vdots & \vdots & \ddots & \vdots\\
\theta_{J,1} & \theta_{J,2} & \ldots & \theta_{J,K} \\
\end{array} \right]
\end{equation*}
which has one row for each respondent and one column for each item, where $\theta_{j,k}$ is the amount that respondent $j$ values item $k$ (or more generally, the amount that respondent $j$ believes item $k$ answers the question being asked).  In the New York City example described above, $\theta_{j,k}$ could be the amount that a specific respondent values the idea ``Open schoolyards across the city as public playgrounds."  

Three features of the response data complicate the process of estimating the opinion matrix $\mbox{\boldmath$\Theta$}$.  First, because the wiki survey is greedy, we have an unequal number of responses from each respondent.  Second, because the wiki survey is collaborative, there are some items that can never be presented to some respondents.  For example, if respondent $j$ contributed an item, then none of the previous respondents could have seen that item.  Collectively, the greediness and the collaborativeness mean that in practice we often have to estimate a respondent's value for an item that she has never encountered.  The third problem is that responses are in the form of pairwise comparisons, which means that we can only observe a respondent's relative preference between two items, not her absolute feeling about either item. 

In order to address these three challenges, we propose a statistical model that assumes that respondents' responses reflect their relative preferences between items (i.e., the Thurstone-Mosteller model~\cite{thurstone_method_1927, mosteller_remarks_1951, stern_continuum_1990}) and that the distribution of preferences across respondents for each item follows a normal distribution; see SI 2 for more information.  Given these assumptions and weakly informative priors, we can perform Bayesian inference to estimate the $\theta_{j,k}$'s that are most consistent with the responses that we observe and the assumptions that we have made.  One important feature of this modeling strategy is that for those who contribute many responses, we can better estimate their row in the opinion matrix, and for those who contribute fewer responses, we have to rely more on the pooling of information from other respondents (i.e., imputation).  The specific functional forms that we assume (see SI 2) result in the following posterior distribution, which resembles a hierarchical probit model:
\begin{eqnarray}
p(\boldsymbol{\theta}, \boldsymbol{\mu} \mid \boldsymbol{Y}, \boldsymbol{X}, \sigma, \boldsymbol{\mu_0}, \boldsymbol{\tau^2_{0}}) & \propto & \prod_{i=1}^V \Phi(\boldsymbol{x}_i^T \boldsymbol{\theta})^{y_i} (1-\Phi(\boldsymbol{x}_i^T \boldsymbol{\theta}))^{1-y_i} \times \prod_{j=1}^J \prod_{k=1}^K N(\theta_{j,k} \mid \mu_{k}, \sigma) \nonumber \\
& & \times \prod_{k=1}^K N(\mu_k \mid \mu_{0[k]}, \tau^2_{0[k]})
\end{eqnarray}
where $\boldsymbol{X}$ is an appropriately constructed design matrix, $\boldsymbol{Y}$ is an appropriately constructed outcome vector, $\boldsymbol{\mu} = \mu_1 \ldots \mu_K$ represents the mean appeal of each item, and $\boldsymbol{\mu_0} = \mu_{0[1]} \ldots \mu_{0[K]}$ and $\boldsymbol{\tau^2_0} = \tau^2_{0[1]} \ldots \tau^2_{0[K]}$ are parameters to the priors for mean appeal of each item ($\boldsymbol{\mu}$).  

This statistical model is just one of many possible approaches to estimating the opinion matrix from the response data, and we hope that future research will develop improved approaches.  In SI 2, we fully derive the model, discuss situations in which our modeling assumptions might not hold, and describe the Gibbs sampling approach that we use to make repeated draws from the posterior distribution. Computer code to make these draws was written in \texttt{R}~\cite{r_core_team_r:_2014} and utilized the following packages: \texttt{plyr}~\cite{wickham_split-apply-combine_2011}, \texttt{multicore}~\cite{urbanek_multicore:_2011}, \texttt{bigmemory}~\cite{kane_bigmemory:_2011}, \texttt{truncnorm}~\cite{trautmann_truncnorm:_2011}, \texttt{testthat}~\cite{wickham_testthat:_2011}, \texttt{Matrix}~\cite{bates_matrix:_2011}, and \texttt{matrixStats}~\cite{bengtsson_matrixstats:_2013}.  

\subsubsection{Summarizing opinion matrix}

Once estimated, the opinion matrix $\mbox{\boldmath$\Theta$}$ may include hundreds of thousands of parameters
---there are often thousands of respondents and hundreds of items---that are measured on a non-intuitive scale.  Therefore, the second step of our analysis is to summarize the opinion matrix $\mbox{\boldmath$\Theta$}$ in order to make it more interpretable.  The ideal summary of the opinion matrix will likely vary from setting to setting, but our preferred summary statistic is what we call the score of each item, $\widehat{s}_i$, which is the estimated chance that it will beat a randomly chosen item for a randomly chosen respondent.  That is,
\begin{equation}
\widehat{s}_i =  \frac{\sum_{j=1}^J \sum_{k \neq i} \Phi(\hat{\theta}_{j,i} - \hat{\theta}_{j,k})}{J \times (K-1)} \times 100
\label{eq:modeled_score}
\end{equation}
The minimum score is 0 for an item that is always expected to lose, and the maximum score is 100 for an item that is always expected to win.  For example, a score of 50 for the idea ``Open schoolyards across the city as public playgrounds'' means that we estimate it is equally likely to win or lose when compared to a randomly selected idea for a randomly selected respondent.  To construct 95\% posterior intervals around the estimated scores, we use the $t$ posterior draws of the opinion matrix ($\mbox{\boldmath$\Theta^{(1)}$}, \mbox{\boldmath$\Theta^{(2)}$}, \ldots, $\mbox{\boldmath$\Theta^{(t)}$}) to calculate $t$ posterior draws of  $\boldsymbol{s}$ ($\widehat{\boldsymbol{s}}^{(1)}, \widehat{\boldsymbol{s}}^{(2)}, \ldots, \widehat{\boldsymbol{s}}^{(t)}$).  From these draws, we calculate the 95\% posterior intervals around $\widehat{s}_i$ by findings values $a$ and $b$ such that $Pr(\widehat{s}_i > a) = 0.025$ and $Pr(\widehat{s}_i < b) = 0.025$~\cite{gelman_bayesian_2003}.

We chose to conduct a two-step analysis process---estimating and then summarizing the opinion matrix, $\mbox{\boldmath$\Theta$}$---rather than estimating the scores directly for three reasons.  First, we believe that making the opinion matrix, $\mbox{\boldmath$\Theta$}$, an explicit target of inference underscores the possible heterogeneity of preferences among respondents.  Second, by estimating the opinion matrix as an intermediate step, our approach can be extended to cases in which co-variates are added at the level of the respondent (e.g., gender, age, income, etc.) or at the level of the item (e.g., about the economy, about the environment, etc.).  Finally, although we are currently most interested in the score as a summary statistic, there are many possible summaries of the opinion matrix that could be important, and by estimating $\mbox{\boldmath$\Theta$}$ we enable future researchers to choose other summaries that may be important in their setting (e.g., which items cluster together such that people who value one item in the cluster tend to value other items in the cluster?).  We return to some possible improvements, extensions, and generalizations in the Discussion.  

\section{Case studies}

To show how pairwise wiki surveys operate in practice, in this section we describe two case studies in which the All Our Ideas platform was used for collecting and prioritizing community ideas for policymaking: New York City's PlaNYC 2030 and the Organisation for Economic Co-operation and Development (OECD)'s ``Raise Your Hand'' initiative.  As described previously, the New York City Mayor's Office conducted a wiki survey in order to integrate residents' ideas into the 2011 update to the City's long-term sustainability plan. The wiki survey asked residents to contribute their ideas about how to create ``a greener, greater New York City" and to vote on the ideas of others.  The OECD's wiki survey was created in preparation for an Education Ministerial Meeting and an Education Policy Forum on ``Investing in Skills for the 21st Century.''  The OECD sought to bring fresh ideas from the public to these events in a democratic, transparent, and bottom-up way by seeking input from education stakeholders located around the globe. To accomplish these goals, the OECD created a wiki survey to allow respondents to contribute and vote on ideas about ``the most important action we need to take in education today.''

We assisted the New York City Mayor's Office and the OECD in the process of setting up their wiki surveys, and spoke with officials of both institutions multiple times over the course of survey administration. We also conducted qualitative interviews with officials from both groups at the conclusion of survey data collection in order to better understand how the wiki surveys worked in practice, contextualize the results, and get a better sense of whether the use of a wiki survey enabled the groups to obtain information that might have been difficult to obtain via other data collection methods.  Unfortunately, logistical considerations prevented either group from using a probabilistic sampling design.  Therefore, we can only draw inferences about respondents, who should not be considered a random sample from some larger population.  However, wiki surveys can be used in conjunction with probabilistic sampling designs, and we will return to the issue of sampling in the Discussion.

\subsection{Quantitative results}

\begin{figure}
  \includegraphics[width=\textwidth]{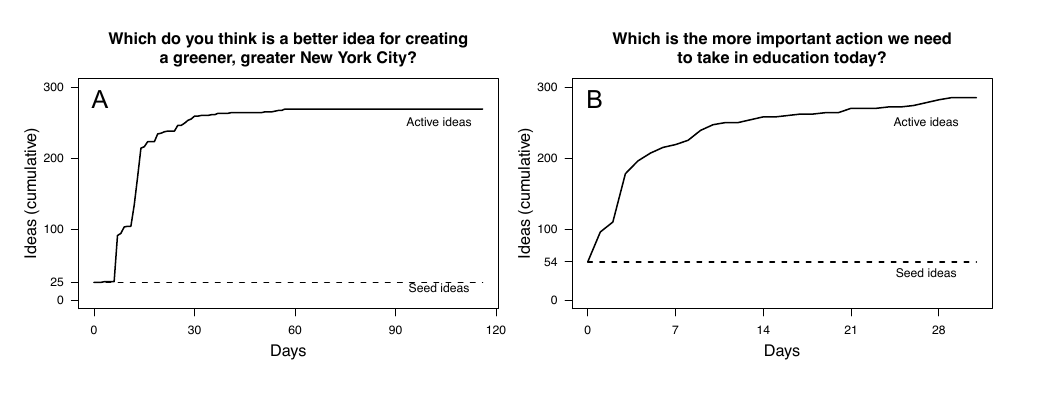}
  \centering
   \caption{{\bf Cumulative number of activated ideas for PlaNYC [A] and OECD [B].} The PlaNYC wiki survey ran from October 7, 2010 to January 30, 2011. The OECD wiki survey ran from September 15, 2010 to October 15, 2010. In both cases the pool of ideas grew over time as respondents contributed to the wiki survey.  PlaNYC had \planycseedideas~seed ideas and \planycuploadedideas~user-contributed ideas, \planycuploadedactiveideas~of which the Mayor's Office activated.  The OECD had \oecdseedideas~seed ideas (\oecdseedideasdeactivated~of which it deactivated during the course of the survey), and \oecduploadedideas~user-contributed ideas, \oecduploadedactiveideas~of which it activated.  In both cases, ideas that were deemed inappropriate or duplicative were not activated.}
   \label{fig:ideas} %
\end{figure}

The pairwise wiki surveys conducted by the New York City Mayor's Office and the OECD had similar patterns of respondent participation.  In the PlaNYC wiki survey, \planycrespondents~respondents contributed \planycresponses~responses, and in the OECD wiki survey \oecdrespondents~respondents contributed \oecdresponses~responses.  Further, respondents contributed a substantial number of new ideas (\planycuploadedideas~for PlaNYC, and \oecduploadedideas~for OECD). Of these contributed ideas, those that the wiki survey creators deemed inappropriate or duplicative were not activated. In the end, the number of ideas under consideration was dramatically expanded.  For PlaNYC the number of active ideas in the wiki survey increased from \planycseedideas~to \planycactiveideas, a 10-fold increase, and for the OECD from \oecdseedideas~to \oecdactiveideas, a 5-fold increase (Fig.~\ref{fig:ideas}).

\begin{figure}
  \includegraphics[width=\textwidth]{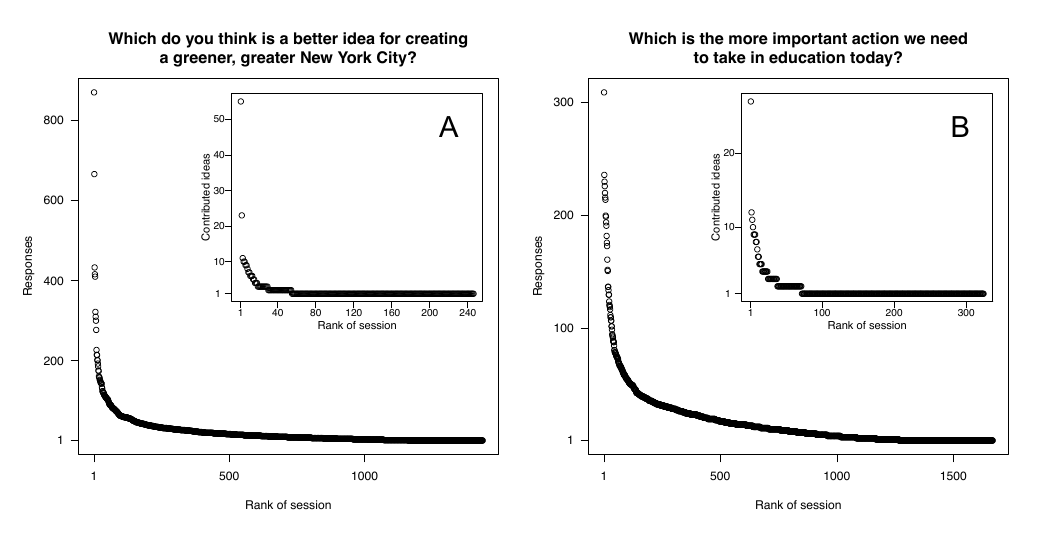}
  \centering
  \caption{{\bf Distribution of contribution per respondent for PlaNYC [A] and OECD [B].}  Both the number of responses per respondent and the number of ideas contributed per respondent show a ``fat head'' and a ``long tail." Note that the scales on the figures are different.}
   \label{fig:user_activity} %
\end{figure}

Within each survey, the level of respondent contribution varied widely, in terms of both number of responses and number of ideas contributed, as we expected given the greedy nature of the wiki survey.   In both cases, the distributions of both responses and contributed ideas contained ``fat heads" and ``long tails" (see Fig.~\ref{fig:user_activity}). If the wiki surveys captured only a fixed amount of information per respondent---as opposed to capturing all levels of effort---a significant amount of information would have been lost.  For instance, if we only accepted the first 10 responses per respondent and discarded all respondents with fewer than 10 responses, approximately 75\% of the responses in each survey would have been discarded.  Further, if we were to limit the number of ideas contributed to one per respondent, as is typical in surveys with one and only one ``other box," we would have excluded a significant number of new ideas: nearly half of the user-contributed ideas in the PlaNYC survey\footnote{In this case, the number of user-contributed ideas per session is somewhat difficult to interpret because some user-contributed ideas were bulk-uploaded by the Mayor's Office following community meetings at which ideas were recorded on paper.  Unfortunately no records were kept of this bulk uploading, so we cannot distinguish it from other respondent behavior.} and about 40\% in the OECD survey.

\begin{figure}
  \includegraphics[width=\textwidth]{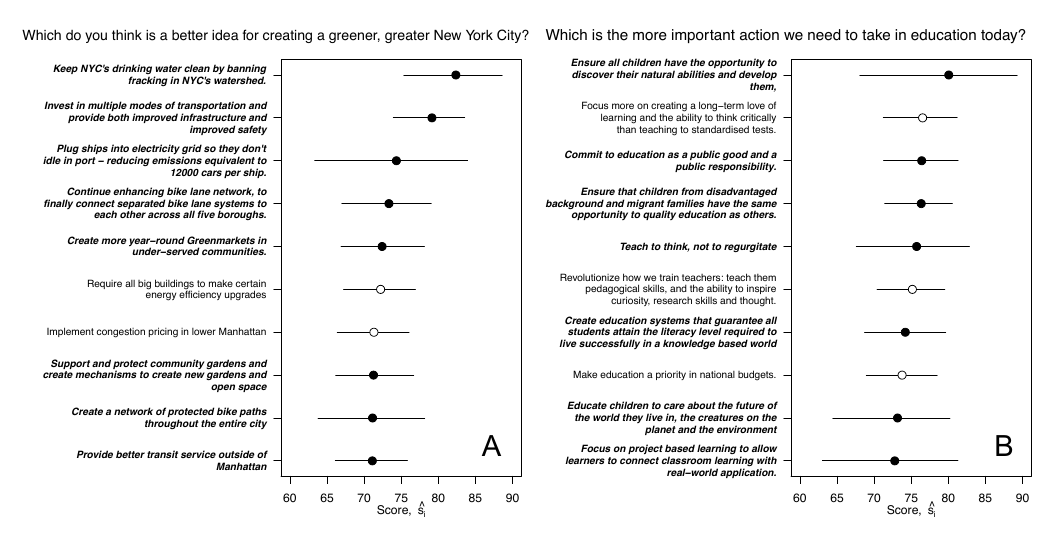}
  \centering
  \caption{{\bf Ten highest-scoring ideas for PlanNYC [A] and OECD [B].}  Ideas that were contributed by respondents are printed in a bold/italic font and marked by closed circles; seed ideas are printed in a standard font and marked by open circles.  In the case of PlaNYC, \planyctoptenuploaded~of the 10 highest-scoring ideas were contributed by respondents.  In the case of the OECD, \oecdtoptenuploaded~of the 10 highest-scoring ideas were contributed by respondents.  Horizontal lines show 95\% posterior intervals.}
     \label{fig:top10} %
\end{figure}

In both cases, many of the highest-scoring ideas were contributed by respondents.  For PlaNYC, \planyctoptenuploaded~of the top 10 ideas were contributed by users, as were \oecdtoptenuploaded~of the top 10 ideas for the OECD (Fig.~\ref{fig:top10}).  These high-scoring user-contributed ideas highlight a strength of pairwise relative wiki surveys relative to both surveys and interviews.  With a survey, it would have been difficult to learn about these new user-contributed ideas, and with an interview it would have been difficult to empirically assess the support that respondents have for them.

\begin{figure}
  \includegraphics[width=\textwidth]{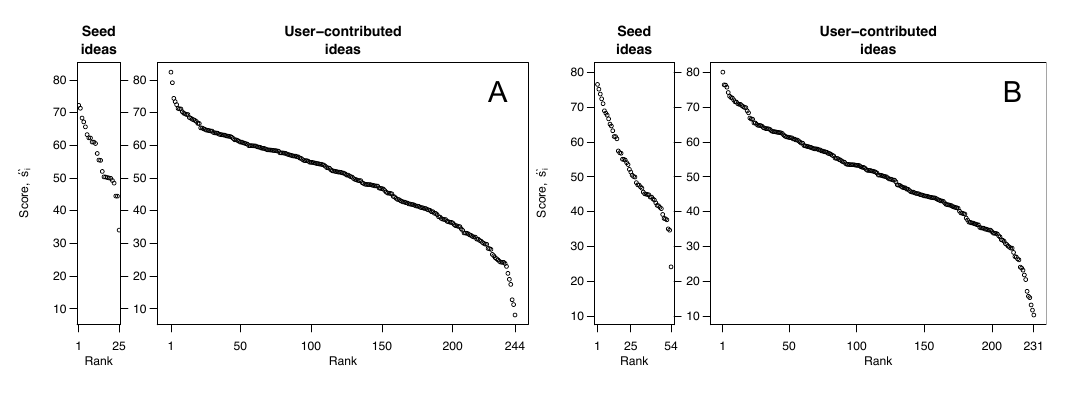}
  \centering
  \caption{{\bf Distribution of scores of seed ideas and user-contributed ideas for PlaNYC [A] and OECD [B].}  In both cases, some of the lowest-scoring ideas were user-contributed, but critically, some of the highest-scoring ideas were also user-contributed.  In general, the large number of user-contributed ideas, combined with their high variance, means that they typically include some extremely popular ideas.  Posterior intervals for each estimate are not shown.}
  \label{fig:compare_seed_uploaded} %
\end{figure}

Building on these specific results, we can begin to formulate a general model that describes the situations in which many of the top scoring items will be contributed by respondents. Three mathematical factors determine the extent to which an idea generation process will produce extreme outcomes (i.e., high scoring ideas): the number of ideas, the mean of ideas' scores, and the variance of ideas' scores~\cite{girotra_idea_2010}.  In both of these case studies, there were many more user-contributed ideas than seed ideas, and they had higher variance in scores (Fig.~\ref{fig:compare_seed_uploaded}).  These two features---volume and variance---ensured that many of the highest-scoring ideas were contributed by respondents, even though these ideas had a lower mean score than the seed ideas.  Thus, in settings in which researchers seek to discover the highest-scoring ideas, the high variance and high volume of user-contributed ideas make them a likely source of these extreme outcomes.

\subsection{Qualitative results}

Because user-contributed ideas that score well are likely to be of interest---in fact, they highlight the value of the collaborativeness of wiki surveys---we sought to understand more about these items by conducting interviews with the creators of the PlaNYC and OECD wiki surveys. Based on these interviews, as well as interviews with six other wiki survey creators, we identified two general categories of high-scoring user-contributed ideas: \emph{novel information}---that is, substantively new ideas that were not anticipated by the wiki survey creators---and \emph{alternative framings}---that is, new and resonant ways of expressing existing ideas.

Some high-scoring user-contributed ideas contained information that was novel to the wiki survey creator.  For example, in the PlaNYC context, the Mayor's Office reported that user-contributed ideas were sometimes able to bridge multiple policy arenas (or ``silos") that might have been more difficult connections to make for office staff working within a specific arena. For instance, consider the high-scoring user-contributed idea ``plug ships into electricity grid so they don't idle in port---reducing emissions equivalent to 12000 cars per ship." The Mayor's Office suggested that staff may not have prioritized such an idea internally (it did not appear on the Mayor's Office's list of seed ideas), even though the idea's high score suggested public support for this policy goal: ``[T]his relates to two areas.  So plugging ships into electricity grid, so that's one, in terms of energy and sourcing energy.  And it relates to freight. [Question: Okay, which are two separate silos?] Correct, so freight is something that we're looking closer at. ... And emissions, reducing emissions, is something that's an overall goal of the plan. ... So this has a lot of value to it for us to learn from" (interview with Ibrahim Abdul-Matin, New York City Mayor's Office, December 12, 2010).

Other user-contributed ideas suggested alternative framings for existing ideas.  For instance, the creators of the OECD wiki survey noted that high-scoring, user-contributed ideas like ``Teach to think, not to regurgitate" ``wouldn't be formulated in such a way [by the OECD]. ... [I]t's very un-OECD-speak, which we liked" (interview with Julie Harris, OECD, February 3, 2011). More generally, OECD staff noted that ``what for me has been most interesting is that ... those top priorities [are] very much couched in the language of principles[. ...] It's sort of constitutional language'' (interview with Joanne Caddy, OECD, February 15, 2011). PlaNYC's wiki survey creators also described the importance of user-contributed ideas being expressed in unexpected ways. The top-scoring idea in PlaNYC's wiki survey, contributed by a respondent, was ``Keep NYC's drinking water clean by banning fracking in NYC's watershed'';\footnote{``Fracking,'' short for hydraulic fracturing, is a drilling technique for extracting oil and natural gas.} Mayor's Office staff indicated that the office would have used more general language about protecting the watershed, rather than referencing fracking explicitly: ``[W]e talk about it differently.  We'll say, `protect the watershed.'  We don't say, `protect the watershed from fracking' " (interview with Ibrahim Abdul-Matin, New York City Mayor's Office, December 12, 2010).

Taken together, these two case studies suggest that pairwise wiki surveys can provide information that is difficult, if not impossible, to gather from more traditional surveys or interviews. This unique information comes from high-scoring user-contributed ideas, and may involve both the content of the ideas and the language used to frame them.

\section{Discussion}

In this paper we propose a new class of data collection instruments called \emph{wiki surveys}.  By combining insights from traditional survey research and projects such as Wikipedia, we propose three general principles that all wiki surveys should satisfy: they should be greedy, collaborative, and adaptive.  Designing an instrument that satisfies those three criteria introduces a number of challenges for data collection and data analysis, which we attempt to resolve in the form of a pairwise wiki survey.  Through two case studies we show that pairwise wiki surveys can enable data collection that would be difficult with other methods.  Moving beyond these proof-of-concept case studies to a fuller understanding of the strengths and weaknesses of pairwise wiki surveys, in particular, and wiki surveys, in general, will require substantial additional research.  

One next step for improving our understanding of the measurement properties of pairwise wiki surveys would be additional studies to assess the consistency and validity of responses.  Consistency could be assessed by measuring the extent to which respondents provide identical responses to the same pair and provide transitive responses to a series of pairs.  Assessing validity would be more difficult, however, because wiki surveys tend to measure subjective states, such as attitudes, for which gold-standard measures rarely exist~\cite{turner_surveying_1984}.  Despite the inherent difficulty of validating measures of subjective states, there are several approaches that could lead to increased confidence in the validity of pairwise wiki surveys~\cite{fowler_improving_1995}.  First, studies could be done to assess discriminant validity by measuring the extent to which groups of respondents who are thought to have different preferences produce different wiki survey results.  Second, construct validity could be assessed by measuring the extent to which responses for items that we believe to be similar are in fact similar.  Third, studies could assess predictive validity by measuring the ability of results from pairwise wiki surveys to predict the future behavior of respondents. Finally, the results of pairwise wiki surveys could be compared to data collected through other quantitative and qualitative methodologies. 

Another area for future research about pairwise wiki surveys is improving the statistical methods used to estimate the opinion matrix---either by choosing pairs more efficiently or developing more flexible statistical models.  First, one could develop algorithms that would choose pairs so as to maximize the amount learned from each respondent. However, maximizing the amount of information per response~\cite{lindley_measure_1956, glickman_adaptive_2005, pfeiffer_adaptive_2012} may not maximize the amount of information per respondent, which is determined by both the information per response and the number of responses provided by the respondent~\cite{von_ahn_designing_2008}.  That is, an algorithm that chooses very informative pairs from a statistical perspective might not be effective if people do not enjoy responding to those kinds of pairs.  Thus, algorithms could be developed to address both maximization of information per pair and to encourage participation by, for example, choosing pairs to which respondents enjoy responding. In addition to choosing pairs more efficiently, we believe that substantial progress can be made by developing more flexible and general statistical models for estimating the opinion matrix from a set of responses.  For example, the statistical model we propose could be extended to include co-variates at the level of the respondent (e.g., age, gender, level of education, etc.) and at the level of the item (e.g., phrase structure, item topic, etc.).  Another modeling improvement would involve creating more flexible assumptions about the distributions of opinions among respondents. These methodological improvements could be assessed by their robustness and their ability to improve the prediction of future responses (e.g.,~\cite{mao_capturing_2013}).

Another important next step is to combine pairwise wiki surveys with probabilistic sampling methods, something that was logistically impossible in our case studies.  If one thinks of survey research as a combination of sampling and interacting with respondents~\cite{conrad_envisioning_2008}, then pairwise wiki surveys should be considered a new way of interacting with respondents, not a new way of sampling.  However, pairwise wiki surveys can be naturally combined with a variety of different sampling designs.  For example, researchers wishing to employ pairwise wiki surveys with a nationally representative sample can make use of commercially available online panels~\cite{baker_research_2010, brick_future_2011}.  Further, researchers wishing to study more specific groups---e.g., workers in a firm or residents in a city---could draw their own probability samples from administrative records. 

Given the significant amount of work that remains to be done, we have taken a number of concrete steps to facilitate the future development of pairwise wiki surveys.  First, we have made it easy for other researchers to create and host their own pairwise wiki surveys at \url{www.allourideas.org}.  Further, the website enables researchers to download detailed data from their survey which can be analyzed in any way that researchers find appropriate. Finally, we have made all of the code that powers \url{www.allourideas.org} available open-source so that anyone can modify and improve it.  We hope that these concrete steps will stimulate the development of pairwise wiki surveys.  Further, we hope that other researchers will create different types of wiki surveys, particularly wiki surveys in which respondents themselves help to generate the questions~\cite{sullivan_alternative_1979, gal_answering_2011}. We expect that the development of wiki surveys will lead to new and powerful forms of open and quantifiable data collection.
 
\newpage

\appendix
\renewcommand{\theequation}{SI 1-\arabic{equation}}
\setcounter{equation}{0}  %
\renewcommand{\thesection}{SI \arabic{section}}
\setcounter{section}{0}  %

\section{Website implementation}

In this SI, we describe the procedures used for data collection on our website, \url{www.allourideas.org}. When implementing pairwise wiki surveys on the website, we encountered three main methodological issues: 1) choosing pairs to present to respondents; 2) using the responses to estimate the score; and 3) ensuring data quality.  In all cases, we solved these problems using relatively simple heuristic approaches. Our heuristic for score estimation on the website differs from the technique used in the data analysis section of the paper; we present it here for completeness.  We are confident that many of these approaches will be improved based on future research.

\subsection{Selection of pairs}
\label{sec:select_pairs}

The simplest way to select pairs for the respondents would be to sample with uniform probability from the set of pairs.  However, because pairwise wiki surveys are collaborative, respondents contribute new items throughout the process, which means that pairs with user-contributed items will tend to have fewer responses.  Therefore, the simple approach would result in more responses---and therefore more precise estimates---for seed items than user-contributed items.  This disparity is problematic because the user-contributed items are potentially the most interesting.  Instead, it is preferable to spread the responses more evenly over the set of pairs. Therefore, we developed a ``catch up'' algorithm, which shows pairs with fewer completed responses with higher probability.  In essence, it helps newer pairs ``catch up'' to older ones in terms of number of responses. Specifically, the draw-wise probability for a given pair $(i,j)$ is:
\begin{equation}
p_{i,j} = \frac{min \left( \frac{ \frac{1}{(n_{i,j}+1)^\alpha} } {c_1}, \tau \right)}{c_2}
\label{eq:throttled_catchup}
\end{equation}
where $n_{i,j}$ is the number of votes on prompt $(i,j)$, $\alpha$ is a parameter that weights the number of responses, and $\tau$ is a ``throttle'' to ensure that the draw-wise probability never exceeds some threshold (it could create a poor user experience if the same pair had a draw-wise probability of, say, 0.5).  Finally, $c_1$ and $c_2$ are normalizing constants to ensure that the distribution sums to 1.\footnote{The normalizing constants are $c_1 = \sum_i \frac{1}{(n_{i,j}+ 1)^\alpha}$ and $c_2= \sum_i min \left( \frac{ \frac{1}{(n_{i,j} + 1)^\alpha}} {c_1}, \tau \right)$ where $\tau$ is the throttle, the maximum probability for a pair appearing in a draw.}  Although somewhat awkward-looking, Eq.~\ref{eq:throttled_catchup} is straightforward to implement and runs very quickly.  As a first step we choose $\alpha=1$ and $\tau=0.05$, but the optimal values of these parameters are an open question.

\subsection{Estimating score}

We decided to make the score of each item available to all respondents in real time.  This requirement for real time calculation made it impossible for us to use the statistical methods described in the data analysis section of this paper.  Therefore, for the website, we developed a simpler method of estimating the score.  In the cases considered in the paper, the two estimates of the score were very similar; there was a correlation of about 0.95 in both cases.  

Recall that the score of an item is the probability that the item will beat a randomly chosen item for a randomly chosen respondent.  Given this focus on the probability of a win, we choose a binomial model.  If one assumes a uniform prior for a binomial random variable, the resulting posterior for the probability of a win follows a Beta distribution~\cite{hoff_first_2009}.  If we multiply the expected value of that Beta distribution by 100 (to place things on a more natural scale), we have
\begin{equation}
\widehat{s}_i^{'}=\frac{(w_i+1)}{(w_i+1) + (l_i+1)} \times 100
\label{eq:simple_score}
\end{equation}
where $w_i$ is the number of wins for item $i$ and $l_i$ is the number of losses for item $i$; see~\cite[Ch. 3]{hoff_first_2009} for a derivation.  Thus, the estimated score ranges from 0 to 100 and resembles a simple winning percentage with an additional term that provides some smoothing. 

This approach is both easy to calculate and reasonably principled because it is derived from standard Bayesian methods.  It also has several desirable practical properties including that it produces a reasonable estimate for new items that have not appeared ($\widehat{s}_i^{'}=50$) and the amount the score changes with any specific vote decreases as the number of votes on the item increase.  However, the approach also has some limitations.  First, it does not account for the fact that responses are nested within respondents.  In other words, a respondent who contributes 100 responses will have 100 times the influence as someone who responds only once.  Also, this approach does not consider the ``strength of schedule'' (i.e., the scores of the items that a given item has competed against).  For example, this scoring approach gives equal weight to an item beating a popular item as to one beating an unpopular item.  For these reasons and others, we developed the model described in the data analysis section of this paper, which does not suffer from these two limitations, but which takes many hours to compute. 

\subsection{Data quality issues}

In all data collection, researchers must be wary of respondents who wish to manipulate results, but those risks are particularly salient in this research.  In order to make our results more manipulation-resistant~\cite{resnick_information_2008}, we flagged some responses as invalid.  These invalid responses were collected, but not included in the final analysis.  There are two ways that a response can be flagged as invalid.  First, if we receive multiple, consecutive responses for the same pair (as would occur if the respondent tried to click several times before the page reloads), then only the first response is marked valid; the others are marked invalid and are not included in the data files generated by \url{www.allourideas.org}.  Second, all responses that occur immediately following the response ``I can't decide'' are marked invalid but are still included in the data files generated by \url{www.allourideas.org}.  These responses are not included in estimation because in a previous pairwise wiki survey we detected a respondent who attempted to manipulate the results by clicking ``I can't decide'' until his or her preferred idea was presented, at which point he or she voted for that idea.  Our flagging procedure prevents this manipulation from influencing the results.  Though our approach probably invalidates some legitimate data, we prefer to err on the side of caution.  Finally, we note that these procedures do not protect against all possible forms of manipulation, and future research will be necessary to make wiki surveys more manipulation-resistant.  In the two case studies presented in this paper, we do not believe that any large-scale manipulations took place.

A second potential data quality issue arose because our pairwise wiki survey website had no login system to verify a respondent's identity.  We decided not to create such a system because we wanted to minimize barriers that might create differential non-response.  However, the lack of authentication means that there is no guarantee that each of our respondents is unique.  Each participant is defined by a ``session'' at the website, and a session is created when a browser that is not currently in a session visits the site.  If there are 10 minutes of inactivity on the site, the current session is terminated; future activity on the site would result in a new session being created. The sessions are tracked with browser cookies.  There are many ways that a single person could create multiple sessions and thus be considered multiple respondents (e.g., by visiting the wiki survey from a new browser or by deleting cookies).  We do not believe that a single person creating multiple sessions caused any large-scale problems in the two case studies presented in this paper. 

\newpage

\renewcommand{\theequation}{SI 2-\arabic{equation}}
\setcounter{equation}{0}  %
\renewcommand{\thesection}{SI \arabic{section}}
\setcounter{section}{1}  %
\renewcommand{\thetable}{SI 2-\arabic{table}}
\setcounter{table}{0}  %
\renewcommand{\thefigure}{SI 2-\arabic{figure}}
\setcounter{figure}{0}  %

\section{Data analysis}

\subsection{Statistical model}

As described in the paper, a main statistical challenge is to use the responses (e.g., Table~\ref{tab:votes}) to estimate the opinion matrix, $\boldsymbol{\Theta}$, which represents how much each respondent values each item.  To do this, we begin by assuming a model for how the votes are generated; a natural first choice would be 
\begin{equation}
Pr(a \mbox{ beats } b \mbox{ in session } j) = F(\theta_{j,a} - \theta_{j,b})
\end{equation}
where $\theta_{j,a}$ is the amount that respondent $j$ values item $a$.  That is, the probability that item $a$ beats item $b$ is a function of the difference in the appeals of the two items $\theta_{j,a}$ and $\theta_{j,b}$.  In previous work, numerous functional forms have been assumed for $F$, but the two common choices are the cumulative standard normal---resulting in the Thurstone-Mosteller model~\cite{thurstone_method_1927, mosteller_remarks_1951}---or the logistic function---leading to the Bradley-Terry model~\cite{bradley_rank_1952}.  In fact, Stern~\cite{stern_continuum_1990} has shown that the Thurstone-Mosteller model and the Bradley-Terry model can both be viewed as special cases of a more general model, and empirically, both models produce estimates that are essentially equivalent~\cite{stern_are_1992}.  However, the Thurstone-Mosteller model is much easier to work with computationally because it facilitates the Gibbs sampling updates as described below.  For that reason we assume that
\begin{equation}
Pr(a \mbox{ beats } b \mbox{ in session } j) = \Phi(\theta_{j,a} - \theta_{j,b})
\label{eq:voting_model}
\end{equation}
where $\Phi$ is the cumulative standard normal distribution.  Thus, we map the difference between the appeals, which ranges from $-\infty$ to $\infty$, to a value that ranges from $0$ to $1$.  Future work could explore the robustness of our estimates to the choice of the standard normal or could attempt to estimate the shape of $F$ directly.  Another extension of the model would allow for ``I can't decide'' responses, which are not included in our current modeling framework.

Given the response model described in equation~\ref{eq:voting_model} and assuming that responses are independent, we can create a design matrix $\boldsymbol{X}$ and outcome vector $\boldsymbol{Y}$ so that the likelihood can be written to resemble a standard probit model,
\begin{equation}
p(\boldsymbol{\theta} \mid \boldsymbol{Y}, \boldsymbol{X}) = \prod_{i=1}^V \Phi(\boldsymbol{x}_i^T \boldsymbol{\theta})^{y_i} (1-\Phi(\boldsymbol{x}_i^T \boldsymbol{\theta}))^{1-y_i}.
\label{eq:model_nohier}
\end{equation}
In this case, $\boldsymbol{X}$ has $V$ rows and $J \times K$ columns, where $V$ is the number of votes, $J$ is the number of respondents, and $K$ is the number of items.  Therefore, $\boldsymbol{x}_i = (x_{i1}, x_{i2}, \ldots x_{im})$ and $m=J \times K$.  In order for the algebra to work out properly, each row in $\boldsymbol{X}$ has a ``1'' in the column of the respondent/item that appeared on the left of the pair and a ``-1'' on the column of the respondent/item that appeared on the right of the pair.  $\boldsymbol{Y}$ is a vector with $V$ entries that has a ``1'' if the item on the left is chosen and ``0'' if the item on the right is chosen.  For example, the votes in Table~\ref{tab:votes} would lead to

\begin{center}
\begin{tabular}{ccc}
$\boldsymbol{Y} = 
\begin{pmatrix} 
\: 1 \: \\ %
0 \\
1 \\
1 \\
0 \\
\end{pmatrix}
$
& and &
$\boldsymbol{X} =$ 
  \begin{blockarray}{cccccccc}
  $\theta_{1,1}$ & $\theta_{1,2}$ & $\theta_{1,3}$ & $\theta_{1,4}$ & $\theta_{2,1}$ & $\theta_{2,2}$ & $\theta_{2,3}$ & $\theta_{2,4}$ \\
    \begin{block}{(cccc|cccc@{\hspace*{5pt}})}
    1 & 0 & 0 & -1 & \BAmulticolumn{4}{c}{\multirow{3}{*}{$0$}}\\
    -1 & 0 & 1 & 0 & &&&&\\
    0 & 0 & -1 & 1 & &&&&\\
    \cline{1-8}%
    \BAmulticolumn{4}{c|}{\multirow{2}{*}{$0$}} & 0 & 0 & 1 & -1\\
    &&& & 0 & -1 & 0 & 1\\
    \end{block}
  \end{blockarray}
\end{tabular}
\end{center}

By explicitly attempting to estimate each respondent's opinion about each item, this modeling approach allows for heterogeneity in the preferences of the respondents.  However, the cost of such flexibility is that there are an enormous number of parameters to be estimated; in each of the case studies in the paper, there were about 375,000 parameters to estimate ($\sim$1,500 respondents $\times$ $\sim$250 items).  Therefore, to add more structure to the problem and to allow for partial pooling of information across respondents~\cite{gelman_data_2006, rossi_bayesian_2006}, we add hierarchical terms in the model that assume that the opinions about each item are normally distributed with an item-specific mean $\mu_k$ and a common standard deviation of $\sigma$ $(\theta_{\cdot, k} \sim N(\mu_k, \sigma))$,  
\begin{equation}
p(\boldsymbol{\theta} \mid \boldsymbol{Y}, \boldsymbol{X}, \boldsymbol{\mu}, \sigma) = 
\prod_{i=1}^V \Phi(\boldsymbol{x}_i^T \boldsymbol{\theta})^{y_i} (1-\Phi(\boldsymbol{x}_i^T \boldsymbol{\theta}))^{1-y_i} 
\times \prod_{j=1}^J \prod_{k=1}^K N(\theta_{j,k} \mid \mu_{k}, \sigma)
\label{eq:hier}
\end{equation}
where $\boldsymbol{\mu} = \mu_1 \ldots \mu_K$ and $\sigma$ is assumed to be 1.  In the case studies considered in this paper, we re-ran the model with $\sigma=0.5$ and $\sigma=2$ as a robustness check, and in both cases, the results were essentially the same as when $\sigma=1$. Future work could improve the model by estimating a $\sigma_k$ for each item or even estimating the functional form that the $\theta_{j,k}$ follow for each $k$.  

\begin{table}
\centering
\begin{tabular}{cccc}
\toprule
Respondent & Response & \multicolumn{2}{c}{Pair} \\
\midrule
1 & 1 & \textbf{[item 1]} & item 4\\
1 & 2 & item 3 & \textbf{[item 1]} \\
1 & 3 & \textbf{[item 4]} & item 3 \\
2 & 4 & \textbf{[item 3]} & item 4 \\ 
2 & 5 & item 4 & \textbf{[item 2]} \\
\bottomrule
\end{tabular}
\caption{{\bf Example responses.}  The bolded item is the one that was chosen by the respondent.}
\label{tab:votes}
\end{table}

Finally we add conjugate priors to yield the following posterior distribution: 
\begin{eqnarray}
p(\boldsymbol{\theta}, \boldsymbol{\mu}, 
\mid \boldsymbol{Y}, \boldsymbol{X}, \sigma, \boldsymbol{\mu_0}, \boldsymbol{\tau^2_{0}}) & \propto &
\prod_{i=1}^V \Phi(\boldsymbol{x}_i^T \boldsymbol{\theta})^{y_i} (1-\Phi(\boldsymbol{x}_i^T \boldsymbol{\theta}))^{1-y_i} \times \prod_{j=1}^J \prod_{k=1}^K N(\theta_{j,k} \mid \mu_{k}, \sigma) \nonumber \\
& & \times \prod_{k=1}^K N(\mu_k \mid \mu_{0[k]}, \tau^2_{0[k]}) 
\label{eq:post}
\end{eqnarray}

\begin{figure}
\centering
\includegraphics[width=0.5\textwidth]{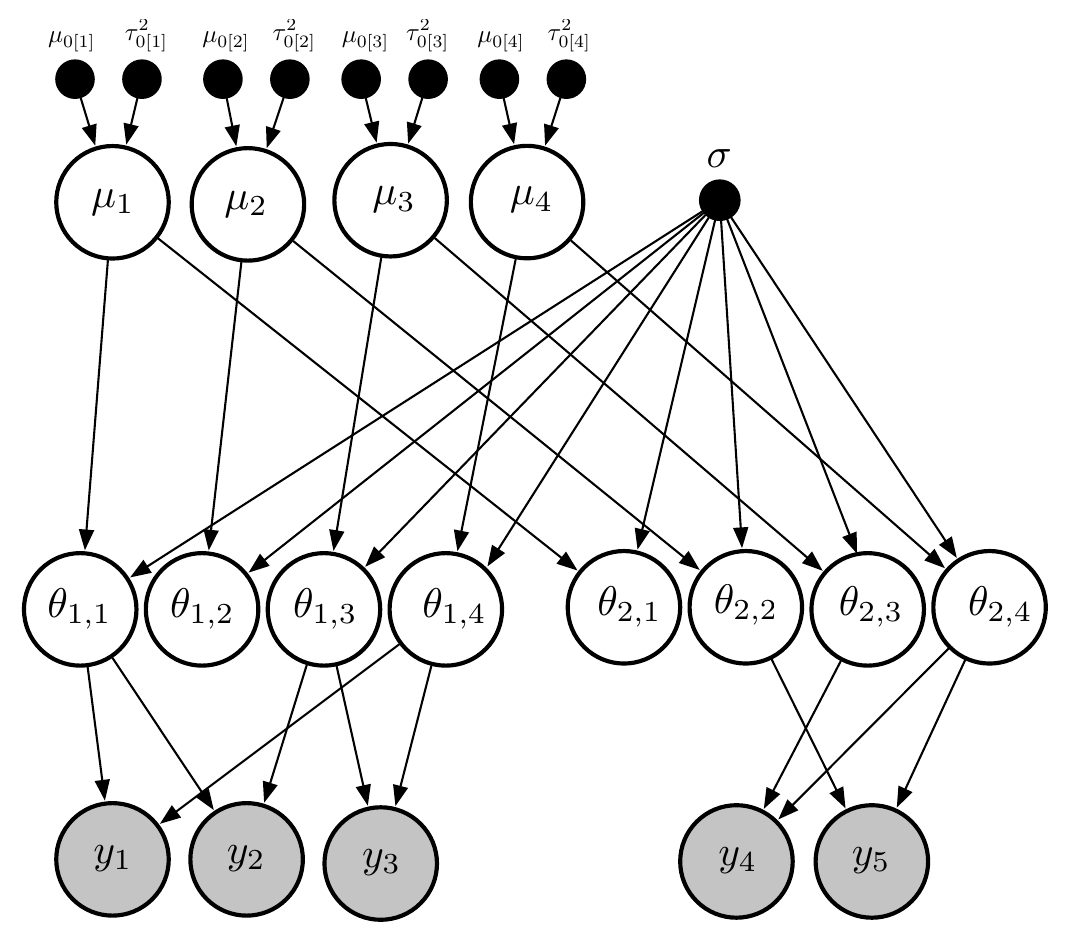}
\caption{{\bf Graphical representation of the model (see Equation~\protect{\ref{eq:post}}).}  This graphical model shows the assumed data generating process for the sample data shown in Table~\ref{tab:votes}.  At the top of the figure, priors are used to generate item-specific means (e.g., $\mu_1$).  Next, these parameters and $\sigma$, which is assumed to be 1, generate the elements of the opinion matrix (e.g., $\theta_{1,1}$).  Finally, these elements of the opinion matrix generate the observed responses (e.g., $y_1$).  The challenge is then to estimate the unknown parameters ($\boldsymbol{\theta}$ and $\boldsymbol{\mu}$) from the observed data.}
\label{fig:graphical}
\end{figure}

As is common in discrete-choice models~\cite{train_discrete_2009}, the model above is only weakly identified because one could add a constant $c$ to all the $\theta$ parameters and leave the posterior largely unchanged (it may be easier to see this non-identifiability from the model for a single response (Eq.~\ref{eq:voting_model})).  Therefore, we pick an arbitrary item to have $\mu_k=0$ which requires setting the hyper-parameters $\mu_{0[k]}=0$ and $\tau_{0[k]}^2=0.000001$.  For the remaining items, we set weakly informative priors: $\mu_{0[k]}=0$, $\tau_{0[k]}^2=4$.  For readers accustomed to graphical models, our model for the 5 responses in Table~\ref{tab:votes} is presented in Figure~\ref{fig:graphical}. 

This model is just one possible model for estimating the opinion matrix from responses.  Further, we do not yet have good procedures for testing modeling assumptions, and we do not know how robust the model is to violations of underlying assumptions.  We suspect that the biggest problems will arise from our assumption about the distribution of opinions across respondents.  In the pairwise wiki surveys analyzed in this paper, it is important to realize that many respondents did not encounter many of the items.  Thus, there are actually two types of $\theta$ parameters, those that are informed by a specific vote ($\boldsymbol{\theta_v}$) and those that are not ($\boldsymbol{\theta_h}$).  We exploit this feature of the data later when describing our approach to computation, but it also has important substantive implications.  Our hierarchical modeling assumption means that we are assuming that the $\theta$'s we estimate based on a specific vote are directly informative of the $\theta$'s for which we have no specific vote (and therefore must make an estimate using data from other respondents).  We can think of two cases in which this assumption might be unreasonable.  First, consider an item uploaded by respondent $j$.  All respondents before $j$ did not have a chance to respond to this item so we will estimate their opinions about the item based on the respondents after $j$.  Therefore, if for some reason the preferences of respondents vary systematically over time, our procedure will not work well.  Second, the greedy nature of the wiki survey could also lead to problems if people who respond many times have systematically different preferences than those who respond fewer times.  For example, imagine that there are two types of people: vegans and non-vegans.  Further, imagine that all vegans love bicycles, all non-vegans hate bicycles, and that vegans contribute more responses than non-vegans.  Now, if we have a respondent $j$ that did not encounter an idea $k$ (``more bike racks in Manhattan''), the model will estimate $\theta_{j,k}$ based on the other $\theta_{\cdot,k} \in \boldsymbol{\theta_v}$.  But, in this case, the $\theta_{\cdot,k} \in \boldsymbol{\theta_v}$ over-represent opinions of vegans relative to non-vegans.  This example shows that an important extension to the model would include co-variates at the level of the respondent and at the level of the item, not only because these are substantively meaningful, but because they can reduce distortions caused by the unequal amount of responses that we have from each respondent.  Diagnostics and robustness will both be important areas of future research for models to estimate the opinion matrix from responses.

\subsection{Computation}

To make draws from this posterior distribution in equation~\ref{eq:full_post} we use Markov chain Monte Carlo, specifically Gibbs sampling~\cite{gelfand_sampling-based_1990}.  That is, we repeatedly draw from the conditional distribution for each parameter given the current values of the other parameters; for a review of Gibbs sampling, see~\cite{gelman_bayesian_2003}. However, 
before attempting to sample from this posterior distribution in this way, we perform two additional steps that greatly facilitate computation, but which do not affect the underlying model that we are estimating. 

First, as described earlier, many respondents do not encounter many of the items.  For example, in the voting data in Table~\ref{tab:votes}, respondent 1 never encountered item 2 and respondent 2 never encountered item 1.  Thus, as described earlier, there are actually two types of $\theta$ parameters, those that are informed by a specific vote (in this case, $\theta_{1,1}$, $\theta_{1,3}$, $\theta_{1,4}$, $\theta_{2,2}$, $\theta_{2,3}$, $\theta_{2,4}$) and those that are not (in this case, $\theta_{1,2}$, $\theta_{2,1}$).  Thus, we note that 
\begin{equation}
p(\boldsymbol{\theta} \mid \boldsymbol{Y}, \boldsymbol{X}, \boldsymbol{\mu}, \sigma) = p(\boldsymbol{\theta_v} \mid \boldsymbol{Y}, \boldsymbol{\dot{X}}, \boldsymbol{\mu}, \sigma) \times p(\boldsymbol{\theta_h} \mid \boldsymbol{\mu}, \sigma)
\end{equation}
where $\boldsymbol{\theta_v}$ are parameters that are estimated from the votes and the hyper-parameters and $\boldsymbol{\theta_h}$ are parameters that depend on the votes only through the hyper-parameters, and $\boldsymbol{\dot{X}}$ is the reduced form of the original design matrix $\boldsymbol{X}$ that only includes columns for $\theta \in \boldsymbol{\theta_v}$.  For example, for the votes in Table~\ref{tab:votes}, $\boldsymbol{\dot{X}}$ is 
\begin{equation*}
\boldsymbol{\dot{X}} = 
\mbox{
  \begin{blockarray}{cccccc}
  $\theta_{1,1}$ & $\theta_{1,3}$ & $\theta_{1,4}$ & $\theta_{2,2}$ & $\theta_{2,3}$ & $\theta_{2,4}$ \\
    \begin{block}{(ccc|ccc@{\hspace*{5pt}})}
    1 & 0 & -1 & \BAmulticolumn{3}{c}{\multirow{3}{*}{$0$}}\\
    -1 & 1 & 0 & &&\\
    0 & -1 & 1  & &&\\
    \cline{1-6}%
    \BAmulticolumn{3}{c|}{\multirow{2}{*}{$0$}} & 0 & 1 & -1\\
    && & -1 & 0 & 1\\
    \end{block}
  \end{blockarray}
}
\end{equation*}

In this simple example $\boldsymbol{\dot{X}}$ is 33\% smaller than $\boldsymbol{X}$, but in both cases considered in the paper the reduction is much more substantial: $\boldsymbol{\dot{X}}$ is about \xdot smaller than $\boldsymbol{X}$.  Reducing the size of the design matrix in this way yields a substantial savings in terms of computing time and RAM needed to make draws from the posterior distribution.   Given this fact, we can re-write equation~\ref{eq:hier} as follows:
\begin{align}
p(\boldsymbol{\theta_v}, \boldsymbol{\theta_h} \mid \boldsymbol{Y}, \boldsymbol{X}, \boldsymbol{\mu}) & \propto & \left( \prod_{i=1}^V \Phi(\boldsymbol{x}_i^T \boldsymbol{\theta_v})^{y_i} (1-\Phi(\boldsymbol{x}_i^T \boldsymbol{\theta_v}))^{1-y_i} \times \prod_{(j,k)}^{\theta_{j,k} \in \boldsymbol{\theta_v}} N(\theta_{j,k} \mid \mu_{k}, \sigma) \right) \nonumber \\ 
 & & \times \left(  1 \times \prod_{(j,k)}^{\theta_{j,k} \in \boldsymbol{\theta_h}} N(\theta_{j,k} \mid \mu_{k}, \sigma)  \right)
\end{align}

A second computational trick is to note that by introducing a latent variable $\boldsymbol{z}$ we are able to sample from the posterior more easily, an approach sometimes called data augmentation.  Roughly, we are assuming that although we observe a discrete outcome $y_i$, there is actually an underlying continuous value $z_i$ that generates $y_i$.  As shown by Albert and Chib \cite{albert_bayesian_1993}, including this continuous latent variable, $z_i$, in our model enables us to sample from the posterior distribution more easily.  For a more thorough discussion of this type of data augmentation, see~\cite{lynch_introduction_2007} and~\cite{jackman_bayesian_2009}.  

Combining these two computational tricks, we are left with the following posterior distribution:
\begin{eqnarray}
 \lefteqn{p(\boldsymbol{\theta_v}, \boldsymbol{\theta_h}, \boldsymbol{z},  \boldsymbol{\mu} \mid \boldsymbol{Y}, \boldsymbol{\dot{X}}, \sigma, \boldsymbol{\mu_0}, \boldsymbol{\tau^2_{0}}) \propto } \nonumber \\
   && \phantom{\times} \left( \prod_{v=1}^V \left( I(z_i > 0)I(y_i=1) + I(z_i < 0)I(y_i=0) \right) \times N(z_i \mid \boldsymbol{\dot{x}_i^T} \boldsymbol{\theta_v}, 1) \times \prod_{(j,k)}^{\theta_{j,k} \in \boldsymbol{\theta_v}} N(\theta_{j,k} \mid \mu_{k}, \sigma) \right)  \nonumber \\
   && \times \left( 1 \times \prod_{(j,k)}^{\theta_{j,k} \in \boldsymbol{\theta_h}} N(\theta_{j,k} \mid \mu_{k}, \sigma)  \right) \times \prod_{k=1}^K N \left( \mu_k \mid \mu_{0[k]}, \tau^2_{0[k]} \right) 
\label{eq:full_post}
\end{eqnarray}

In order to sample from the posterior distribution, we ran three parallel chains from over-dispersed starting values for 200,000 steps, saving every 200th draw, and discarded the first half of each chain as burn-in.  At that point, all parameter estimates had approximately converged, $\hat{R} < 1.1$~\cite{gelman_bayesian_2003}, and so we combined the post burn-in draws to summarize the posterior distribution~\cite{gelman_inference_2011}.  Overall, these computations took about 36 hours per dataset on a fast desktop computer. 

The votes and ideas were then used to fit the model in equation~\ref{eq:full_post} using Gibbs sampling with four update steps.

$\bullet$ Step 1: Draw $\boldsymbol{z}  \mid \boldsymbol{Y}, \boldsymbol{\theta_v}, \boldsymbol{\dot{X}}$

Recall that $\boldsymbol{z}$ is the underlying latent outcome that we cannot observe. Based on ideas developed by Albert and Chib~\cite{albert_bayesian_1993}, we sample $\boldsymbol{z}$ from a truncated normal distribution such that $z_i > 0$ if $y_i=1$ and $z_i < 0$ if $y_i=0$.   That is,
\begin{equation}
{z_i} \sim 
\begin{cases}
N(\boldsymbol{\dot{x}}_i^T \boldsymbol{\theta_v}, 1) I(z_i^* > 0) & \mbox {if } y_i = 1\\
N(\boldsymbol{\dot{x}}_i^T \boldsymbol{\theta_v}, 1) I(z_i^* < 0) & \mbox{if } y_i = 0\\
\end{cases}
\label{eq:truncated_normal}
\end{equation}
where $I$ is an indication function which equals 1 when its argument is true and 0 when false~\cite{jackman_bayesian_2009}.  This indicator function ensures that we are drawing from a properly truncated distribution.  Computationally, we draw from the truncated normal using the \texttt{truncnorm} package in \texttt{R}~\cite{trautmann_truncnorm:_2011}.

$\bullet$ Step 2: Draw $\boldsymbol{\theta_v} \mid \boldsymbol{z}, \boldsymbol{\mu}, \boldsymbol{\dot{X}}, \sigma$

Under the data augmentation we used~\cite{albert_bayesian_1993}, once we have simulated $\boldsymbol{z}$, the latent outcome, we are left with a standard hierarchical linear model.  To update $\boldsymbol{\theta_v}$ we use the ``all-at-once'' approach described in Gelman et al.~\cite{gelman_using_2008}.

That is,
\begin{equation}
\boldsymbol{\theta_v} \sim N(\boldsymbol{\hat{\theta}_d}, \boldsymbol{\hat{V}_{\theta_v}})
\end{equation}
where

\begin{center}
\begin{tabular}{ccc}
$\boldsymbol{\hat{\theta}_d} = (\boldsymbol{\widetilde{X}}^{T} \boldsymbol{\widetilde{\Sigma}}^{-1} \boldsymbol{\widetilde{X}})^{-1} \boldsymbol{\widetilde{X}}^T \boldsymbol{\widetilde{\Sigma}}^{-1} \boldsymbol{\widetilde{Y}}$
&, & 
$\boldsymbol{\hat{V}_{\theta_v}} = ( \boldsymbol{\widetilde{X}}^{T} \boldsymbol{\widetilde{\Sigma}}^{-1} \boldsymbol{\widetilde{X}})^{-1}$
\end{tabular}
\end{center}

\begin{center}
\begin{tabular}{ccccc}
$\boldsymbol{\widetilde{X}} = \left (
   \begin{array}{c}
   \boldsymbol{\dot{X}} \\
   \boldsymbol{I}
   \end{array}
   \right )
$
&, & 
$
\boldsymbol{\widetilde{Y}} = \left (
   \begin{array}{c}
   \boldsymbol{Y} \\
   \boldsymbol{\tilde{\mu}}\\
   \end{array}
   \right )
$
&, and &
$
\boldsymbol{\widetilde{\Sigma}} = \left (
   \begin{array}{cc}
   \boldsymbol{\Sigma_y} & \boldsymbol{0} \\
   \boldsymbol{0} & \boldsymbol{\Sigma_\theta} \\
   \end{array}
   \right ).
$
\end{tabular}
\end{center}

Further, $\boldsymbol{I}$ is the identity matrix, $\boldsymbol{\Sigma_y} = \mbox{Diag}(1)$, $\boldsymbol{\Sigma_\theta} = \mbox{Diag}(\sigma)$, and $\boldsymbol{\widetilde{\mu}}$ is a vector that is the same length as $\boldsymbol{\theta_v}$ and represents an ``expanded'' version of $\boldsymbol{\mu}$.  That is, if the $i^{th}$ column of $\boldsymbol{\dot{X}}$ represents item $k$ (independent of what respondent is involved), then the $i^{th}$ element of $\boldsymbol{\widetilde{\mu}}$ is $\mu_k$.

Computationally, we note that $\boldsymbol{\widetilde{X}}$ and $\boldsymbol{\widetilde{\Sigma}}$ are almost all zeros, so the calculations described above to make a draw are made using sparse matrix routines that are implemented in the \texttt{Matrix} package in \texttt{R}~\cite{bates_matrix:_2011}.  

$\bullet$ Step 3: Update $\boldsymbol{\theta_h} \mid \boldsymbol{\mu}, \sigma$

A large number of the $\theta$ parameters are determined by data only through the hyper-parameters.  For these $\theta$, which we call $\boldsymbol{\theta_h}$, we update as follows:
\begin{equation}
\theta_{j,k} \sim N(\mu_k, \sigma) \quad \forall \quad \theta_{j,k} \in \boldsymbol{\theta_h}
\end{equation}
Thus, this step is roughly like an imputation based on the overall estimated characteristics of the population.  Computationally, no special steps are required to make these updates.

$\bullet$  \textbf{Step 4:} Update $\boldsymbol{\mu} \mid \boldsymbol{\theta_v}, \boldsymbol{\theta_h}, \sigma, \boldsymbol{\mu_0}, \boldsymbol{\tau^2_0}$

\begin{equation}
\mu_k \sim N(\mu, \tau^2)
\end{equation}
where 
\begin{center}
\begin{tabular}{ccc}
$\mu = \frac{ \frac{1}{\tau^2_0} \mu_0+ \frac{n}{\sigma^2} \bar{\theta}_{\cdot, k} }{ \frac{1}{\tau^2_0} + \frac{n}{\sigma^2} }$
& and &
$\tau^2 = \frac{ 1 }{ \frac{1}{\tau^2_0} + \frac{n}{\sigma^2} }$
\end{tabular}
\end{center}
where $\bar{\theta}_{\cdot, k}$ is the mean of the $\theta$ for a specific item $k$ (that is, $\frac{1}{J} \sum_{j=1}^J \theta_{j,k}$) and $n$ is the number of estimates involved (in this case, the number of user-sessions, $J$).  See~\cite[Ch. 6]{hoff_first_2009} for a derivation.  No special computational issues are involved in this update.

\subsection{Data processing}

In order to analyze the data collected in these two case studies, we followed a three-step procedure.  First, using the standard features available to any wiki survey creator at \url{www.allourideas.org}, we downloaded comma-separated value (csv) files that record respondent activity in that wiki survey.  Second, we cleaned the csv files to correct for website errors that occurred during data collection.  More specifically, there were two main data cleaning steps caused by website errors: 1) for a small fraction of participants, \url{www.allourideas.org} automatically created a new session after each vote; and 2) for participants whose sessions timed out after 10 minutes (see SI 1, Sec 3 for more on sessions and session time-outs),  \url{www.allourideas.org} improperly assigned some information to the old session instead of the new session.  After these website issues were discovered while writing this paper, we have improved the code at \url{www.allourideas.org} so that these problems no longer occur.  Finally, after cleaning, we subset the data so that we only estimated parameters for items with at least one win from a valid vote and at least one loss from a valid vote.  For completeness, we describe, in detail, the changes that took place between the data we downloaded from the website and the data that we used for estimation.  

\subsubsection{PlaNYC}

The raw data files from the website included 489 ideas---25 seed ideas and 464 user-contributed ideas---as well as 31,893 responses---26,727 valid votes, 1,988 invalid votes, and 3,178 skips---from 2,094 sessions.  There were no responses or ideas uploaded outside of the appropriate time window of 2010-10-07 to 2011-01-30.  

Cleaning the files caused two main changes.  First, because of participants who were erroneously placed in a new session after each vote, 52 actual sessions had originally been misrepresented as 710 sessions.  Therefore, we collapsed these 710 sessions to the appropriate 52 sessions.  This change in session definitions created 30 valid votes that were immediately preceded by a skip, so we invalidated these 30 votes (see SI 1, Sec 3 for more on invalidating votes after skips to improve data quality). Then, after correcting for errors caused by session time-outs, 9 valid votes were immediately preceded by skips, so we invalidated these 9 votes.  Thus, after cleaning, the files contained 489 ideas---25 seed ideas and 464 user-contributed ideas---as well as 31,893 responses---26,688 valid votes, 2,027 invalid votes, and 3,178 skips---from 1,436 sessions.

When applying our model to the case studies in this paper, we estimated parameters for all items that were active on the final day and had at least one valid win and at least one valid loss.  In PlaNYC there were 269 such items with 26,604 valid votes among them, cast from 1,397 sessions.  Thus, for PlaNYC the opinion matrix, $\boldsymbol{\theta}$, had dimension 1,397 $\times$ 269.

\subsubsection{OECD}

The raw data files from the website included 594 ideas---35 seed ideas and 559 user-contributed ideas---as well as 30,763 responses---27,133 valid votes, 1,338 invalid votes, and 2,292 skips---from 3,373 sessions.  The OECD conducted a period of internal pilot testing from 2010-09-03 to 2010-09-15, and we dropped the 1,747 votes and 164 skips contributed from 182 sessions during this time.  We also converted the 25 ideas contributed during the internal pilot testing to seed ideas.  No responses or ideas were contributed after 2010-10-15.  

Cleaning the files caused two main changes.  First, because of participants who were erroneously placed in a new session after each vote, 104 actual sessions had originally been represented as 1,627 sessions.  Therefore, we collapsed these 1,627 sessions to the appropriate 104 sessions.  This change in session definitions created 93 valid votes that were immediately preceded by a skip, so we invalidated these 93 votes.  Then, after correcting for errors caused by session time-outs, 8 valid votes were immediately preceded by skips, so we invalidated these 8 votes.  Thus, after cleaning, the files contained 594 ideas---60 seed ideas and 534 user-contributed ideas---as well as 28,852 responses---25,393 valid votes, 1,331 invalid votes, and 2,128 skips---from 1,668 sessions.

When applying our model to the case studies in this paper, we estimated parameters for all items that were active on the final day and had at least one valid win and at least one valid loss.  In OECD there were 285 such items with 23,845 valid votes among them, cast from 1,620 sessions.  Thus, for OECD the opinion matrix, $\boldsymbol{\theta}$, had dimension 1,620 $\times$ 285.

\newpage

\bibliography{wikisurvey}

\end{document}